\documentclass[11pt,letterpaper]{article}
\usepackage{graphicx}% Include figure files
\usepackage{bm}% bold math
\usepackage{amssymb}
\usepackage{color}
\usepackage{amsmath}
\usepackage[numbers]{natbib}
\usepackage{enumerate}
\usepackage{amstext}
\usepackage{footmisc}
\usepackage{latexsym}
\usepackage[usenames,dvipsnames]{xcolor}
\usepackage[colorlinks=true,citecolor=Cerulean,linkcolor=RubineRed,urlcolor=Cerulean]{hyperref}
\usepackage{multibib}
\usepackage{ fixfoot}
\usepackage{fancyhdr}
\newcommand{\pas}[1]{\left[#1 \right]}

\newcommand{\ket}[1]{ \left| #1 \rangle\right.}
\newcommand{\bra}[1]{  \left.\langle #1  \right|}

\begin{document}
\title{{\bf Quantum information scrambling  in adiabatically-driven critical systems}}
\author{Ricardo Puebla\href{https://orcid.org/0000-0002-1243-0839}{\includegraphics[scale=0.45]{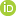}}\,$^{1,\dagger}$, Fernando J. G\'omez-Ruiz\href{https://orcid.org/0000-0002-1855-0671}{\includegraphics[scale=0.45]{orcid}}\,$^{2,*}$}

\date{}
\maketitle
\vspace{-1cm}
\begin{center}
$^{1}${\it Departamento de F{\'i}sica, Universidad Carlos III de Madrid, Avda. de la Universidad 30, 28911 Legan{\'e}s, Spain}\\
$^{2}${\it Departamento de F{\'i}sica Te{\'o}rica, At\'omica y \'Optica, Universidad de Valladolid, 47011 Valladolid, Spain}\\
{\it Correspondence:} $^{\dagger}$\href{maito: rpuebla@fis.uc3m.es}{rpuebla@fis.uc3m.es} and $^{*}$\href{maito:fernandojavier.gomez@uva.es}{fernandojavier.gomez@uva.es}
\end{center}
\begin{abstract}
Quantum information scrambling refers to the spread of the initially stored information over many degrees of freedom of a quantum many-body system. Information scrambling is intimately linked to the thermalization of isolated quantum many-body systems, and has been typically studied in a sudden quench scenario. Here, we extend the notion of quantum information scrambling to critical quantum many-body systems undergoing an adiabatic evolution. In particular, we analyze how the symmetry-breaking information of an initial state is scrambled in adiabatically driven integrable systems, such as the Lipkin--Meshkov--Glick and quantum Rabi models. Following a time-dependent protocol that drives the system from symmetry-breaking to a normal phase, we show how the initial information is scrambled, even for perfect adiabatic evolutions, as indicated by the expectation value of a suitable observable.   We detail the underlying mechanism for quantum information scrambling, its relation to ground- and excited-state quantum phase transitions, and quantify the degree of scrambling in terms of the number of eigenstates that participate in the encoding of the initial symmetry-breaking information. While the energy of the final state remains unaltered in an adiabatic protocol, the relative phases among eigenstates are scrambled, and so is the symmetry-breaking information. We show that a potential information retrieval, following a time-reversed protocol, is hindered by small perturbations, as indicated by a vanishingly small Loschmidt echo and out-of-time-ordered correlators. The reported phenomenon is amenable for its experimental verification, and may help in the understanding of information scrambling in critical quantum many-body systems.
\end{abstract}
\section{Introduction}\label{s:intro}
The process by which quantum information, initially localized within the degrees of freedom of a many-body system, propagates and becomes distributed across the entire system degrees of freedom, is known as information scrambling~\cite{Swingle:16, Swingle:18, Xu:24}. This phenomenon is particularly intriguing from a fundamental perspective, as it delves into the mechanisms of information dispersal and entanglement in complex quantum systems~\cite{Deutsch:91, Touil:21, Touil:24}. Understanding how this scrambling occurs can shed light on the underlying principles governing quantum dynamics, and may have significant implications in quantum computing~\cite{Nielsen}, quantum chaos~\cite{Mezei:17}, quantum thermodynamics~\cite{Chenu:19, Campisi:17, Deffner, Touil:24}, and high energy physics~\cite{Maldacena:16}. 

Quantum many-body systems offer an unrivaled testbed platform to explore the rich interplay between entanglement, thermalization, non-equilibrium dynamics, and even quantum phase transitions (QPTs)~\cite{Polkovnikov:11, Eisert:15, Sachdev}. Quantum information scrambling has been primarily linked with thermalization in closed systems and, thus, it is intimately related to chaotic behavior in quantum many-body systems~\cite{Polkovnikov:11, Srednicki:94, Kaufman:16, Rigol:08, Chenu:19, Mi:21, Zhu:22, Gomez-Ruiz:18}. The complex nature of correlations in non-integrable systems reveals itself in signatures of information scrambling~\cite{Landsman:19}, typically measured in terms of out-of-time-ordered correlation (OTOC) functions~%\FG
{\cite{Maldacena2016}, tripartite mutual information~\cite{Lo_Monaco_2023,Seshadri_PRE18},} and the Loschmidt echo~\cite{Swingle:16, Swingle:18, Xu:24}. Such scrambling of information has been extensively studied in a sudden-quench scenario~\cite{Garttner:17, Zhang:19, Huang:19, Alba:19, Mi:21, Zhu:22, Wang:22, Omanakuttan:23}, where an initial state evolves under its governing and constant Hamiltonian. The integrability of the system has a profound impact on how the information becomes scrambled, with non-integrable systems being better scramblers, as evidenced by the dynamics of OTOCs and the Loschmidt echo (see, for example,~\cite{Alba:19}). %\FG
{Additionally, OTOCs can be utilized to dynamically detect both equilibrium and nonequilibrium phase transitions in various many-body quantum systems~\cite{Heyl_PRL18,Gomez-Ruiz:18,Sun2020,Wei_PRB19,Duan_PRL19}.  However,} the role of time-dependent quenches in quantum information scrambling remains largely unexplored, and even more so when involving critical features of a quantum many-body system. 

In this work, we aim to close this gap and analyze quantum information scrambling in critical systems undergoing a fully coherent adiabatic protocol. %\FG
{The adiabatic protocol ensures zero irreversible work, which may otherwise be significant in critical systems~\cite{Dorner:12,Varizi:20,Bayat:16,Abah:22,Vijayan:23}}. The considered systems feature a normal and symmetry-breaking phase where a $\mathbb{Z}_2$ parity is spontaneously broken. The initial information consists of one of the two possible symmetry-breaking choices, and is encoded in an initial state that populates a large collection of degenerate energy eigenstates that break the discrete parity symmetry of the governing Hamiltonian.  Such symmetry-breaking information remains unaltered under the dynamics of the initial Hamiltonian. This is possible for systems where a symmetry-breaking phase is not only restricted to the ground and first excited states, but rather propagates up to a certain critical excitation energy~\cite{Cejnar:06, Caprio:08, Brandes:13, Stransky:14, Stransky:15, Cejnar:21, Puebla:13, Puebla:13b, Puebla:15, Corps:21, Corps:22b, Corps:23, Gomez-Ruiz:18b, Gomez-Ruiz:16}. Indeed, this is the case for many-body systems with few effective and collective degrees of freedom, such as the Lipkin--Meshkov--Glick (LMG) model~\cite{Lipkin:65, Dusuel:04, Leyvraz:05, Vidal:07, Ribeiro:07, Ribeiro:08}, Dicke model~\cite{Dicke:54, Emary:03}, or quantum Rabi model (QRM)~\cite{Hwang:15, Puebla:16, Bakemeier:12, Puebla:20b, Felicetti:20}. These systems feature a ground-state mean-field quantum phase transition, as well as an excited-state quantum phase transition precisely at a critical excitation energy that divides the spectrum between normal (energy eigenstates with well-defined parity) and a degenerate phase, where eigenstates with opposite parities are degenerated by pairs, as described by a standard double-well semiclassical potential. We stress that the LMG and  QRM are integrable systems, while the Dicke model exhibits a region with chaotic behavior at sufficiently high excitation energies~\cite{Bastarrachea:14, Relano:16, Lobez:16, Corps:22}. It is worth remarking that the paradigmatic one-dimensional Ising model with a transverse field does not meet the previous criterion, since only the two lowest-energy eigenstates are degenerate in the antiferro- or ferromagnetic phase separated by the quantum phase transition~\cite{Sachdev}.

The adiabatic protocol brings the initial symmetry-breaking state towards the normal and back to the degenerate phase by quenching the control parameter of the system. However, while the populations of the energy eigenstates remain constant under the adiabatic theorem, the symmetry-breaking information is scrambled, due to the different and non-commensurable relative phases gained during the protocol, provided that the system is driven into the normal phase. Such dynamical phases are uniformly distributed, resulting in \textit{{adiabatic quantum information scrambling}
} (AQIS). The effectiveness of AQIS is illustrated in the LMG model, and quantified in terms of the final expectation value of a suitable symmetry-breaking observables and the number of populated eigenstates,  as well as with the OTOCs. The robustness of AQIS is analyzed by means of the Loschmidt echo between an adiabatically evolved state and its time-reversed protocol with a small perturbation. AQIS is illustrated for different initial states and another integrable quantum critical model, namely  QRM~\cite{Hwang:15}. 

The article is organized as follows. In Section~\ref{s:AQIS} we present the mechanism that leads to the quantum scrambling of symmetry-breaking information for an adiabatically driven quantum many-body system, i.e., AQIS. In Section~\ref{s:LMG} we first introduce the LMG model, and then present numerical results of the adiabatic quantum information scrambling. A~different model,  QRM, is analyzed in Section~\ref{s:QRM}, and presents similar numerical results, supporting quantum information scrambling. Finally, in Section~\ref{s:conc}, we summarize the main results of the article.

\section{Adiabatic Quantum Information Scrambling}\label{s:AQIS}
Let us start by considering a quantum critical system described by a Hamiltonian $\hat{H}(g)$ that depends on an external controllable parameter $g$, such that
\begin{align}\label{eq:Hg}
    \hat{H}(g)=\sum_{k=0} E_k(g)\ket{\phi_k(g)}\bra{\phi_k(g)},
\end{align}
where $E_k(g)$ denotes the energy of the $k$th eigenstate $\ket{\phi_k(g)}$ for $g$. In addition, we assume that the Hamiltonian commutes with a discrete parity operator $\hat{\Pi}$, $[\hat{H}(g),\hat{\Pi}]=0$, so that the eigenstates can be labeled in terms of the eigenvalues of $\hat{\Pi}$, $+1$ or $-1$, owing to a $\mathbb{Z}_2$ symmetry, i.e., the eigenstates can be written as $\ket{\phi_{k,\pm}(g)}$ with $\hat{\Pi}\ket{\phi_{k,\pm}(g)}=\pm 1 \ket{\phi_{k,\pm}(g)}$, and similarly for its energy $E_{k,\pm }(g)$, so that
\begin{align}\label{eq:Hg_p}
    \hat{H}(g)=\sum_{\substack{k=0\\p=\pm}} E_{k,p}(g)\ket{\phi_{k,p}(g)}\bra{\phi_{k,p}(g)}.
\end{align}

The phase diagram of the Hamiltonian can be split in two regions, namely the normal phase where $E_{k,+}(g)\neq E_{k,-}(g)$, and a symmetry-breaking phase where the eigenstates belonging to a different parity subspace are degenerate, i.e., $E_{k,+}(g)=E_{k,-}(g)$, possibly up to a certain critical excitation energy $E_c(g)$. These two phases are separated by ground- and excited-state quantum phase transitions~\cite{Sachdev, Cejnar:06, Caprio:08, Brandes:13, Stransky:14, Stransky:15, Cejnar:21, Puebla:13, Puebla:13b, Puebla:15, Corps:21}, taking the place of %EE: Please check intended meaning has been retained.
 a critical value $g_c$ and at an excitation energy $E_c(g)$, respectively. Note that in the symmetry-breaking phase, any state of the form $\ket{\phi(g)}=\alpha \ket{\phi_{k,+}(g)}+\beta\ket{\phi_{k,-}(g)}$ with $|\alpha|^2+|\beta|^2=1$ and $E_{k,+}(g)=E_{k,-}(g)$ is also a valid eigenstate of $\hat{H}(g)$. However, for any $\alpha$ such that $|\alpha|\neq 1,0$, $\ket{\phi(g)}$ is no longer an eigenstate of the parity operator, $\hat{\Pi}\ket{\phi(g)}\neq \pm \ket{\phi(g)}$, and thus the symmetry may be spontaneously broken. This effect can be quantified employing the expectation value of a suitable operator $\hat{\mathcal{O}}$, such that  $\bra{\phi_{k,+}(g_0)}\hat{\mathcal{O}}\ket{\phi_{k,-}(g_0)}\neq 0$ and $\bra{\phi_{k,p}(g_0)}\hat{\mathcal{O}}\ket{\phi_{k,p}(g_0)}=0$. Indeed, its expectation value behaves as an order parameter, i.e., $\langle \hat{\mathcal{O}}\rangle=0$ for a symmetric state, and $\langle \hat{\mathcal{O}}\rangle\neq 0$ for symmetry-breaking states, as is the case for magnetization in a standard ferro-to-paramagnetic phase transition. 

We consider an initial state, $\ket{\psi_0}$, which breaks this symmetry for an initial value $g_0$ of the control parameter in the symmetry-breaking phase, i.e., $\bra{\psi_0}\hat{\mathcal{O}}\ket{\psi_0}\neq 0$. This is the information we will consider throughout the rest of the article. Note that, in general, $[\hat{H}(g),\hat{\mathcal{O}}]\neq 0$. Such a state can be written in the eigenbasis of $\hat{H}(g_0)$ as
\begin{align}
    \ket{\psi_0}=\sum_{\substack{k=0\\p=\pm}} c_{k,p}\ket{\phi_{k,p}(g_0)}.
\end{align}

The coefficients, $c_{k,p}$, therefore encode the choice of the symmetry-breaking state, and thus the initial information. Moreover, since $\hat{\mathcal{O}}=\sum_n \gamma_n \ket{\gamma_n}\bra{\gamma_n}$, we can characterize the probability distribution of measuring $\gamma_n$'s in the initial state $\ket{\psi_0}$, which reads as
\begin{align}
    P(\gamma_n)=|\langle \psi_0\ket{\gamma_n}|^2=\Bigg\lvert\sum_{k,p=\pm}c_{k,p}^*\langle \phi_{k,p}(g_0)\ket{\gamma_n}\Bigg\rvert^2.
\end{align}

For a symmetric state, $P(\gamma_n)=P(-\gamma_n)$, and thus $\langle \hat{\mathcal{O}}\rangle=0$. However, for maximally symmetry-broken states, the distribution is only non-zero for one of the two branches, i.e., either $P(\gamma>0)\neq $ or $P(\gamma<0)\neq 0$. For simplicity, and without loss of generality, we will consider initial maximally symmetry-broken states $\ket{\psi_0}$ where $P(\gamma>0)\neq 0$ and $P(\gamma<0)=0$.  
In addition, note that the energy probability distribution of the initial state is simply given by $P(E_k(g))=|c_{k,+}|^2+|c_{k,-}|^2$, since we consider the initial state to be in the symmetry-breaking phase. Therefore, $P(E)$ is independent of the initial information, and thus the energy probability distribution does not depend on how the symmetry is broken. 

We then consider a time-dependent protocol, $g(t)$, that drives the system from $g(0)=g_0$ to $g(\tau)=g_1$, and back to the initial control value, $g(2\tau)=g_0$. For simplicity, we consider a linear ramp, namely 
\begin{equation}\label{eq:gt}
    g(t)=\begin{cases}
    g_0+(g_1-g_0)\frac{t}{\tau}, & \text{for }0\leq t\leq \tau,\\[
1ex]
    g_1+(g_0-g_1)\frac{(t-\tau)}{\tau}, & \text{for }\tau \leq t\leq 2\tau.
    \end{cases}    
\end{equation}

We also assume that the driving is performed very slowly, allowing us to resort to the adiabatic approximation for the final state, $\ket{\psi(2\tau)}$. Indeed, one can write
    \begin{align}\label{eq:PsiAd}
\ket{\psi(2\tau)}\approx \sum_{k,p} c_{k,p} e^{-i\varphi_{k,p}} \ket{\phi_{k,p}(g_0)},
  \end{align}
where $\varphi_{k,p}$ denotes the accumulated phase for the $k$-th eigenstate of parity $p$ upon completing the cycle from $g_0$ to $g_1$ and back to $g_0$, i.e.,
\begin{align}
    \varphi_{k,p}=\frac{2\tau}{(g_1-g_0)}\int_{g_0}^{g_1}dg E_{k,p}(g).
\end{align} 

Under the adiabatic approximation, neither the average energy nor the energy probability distribution $P(E)$ are altered. The relative phases between degenerate eigenstates are irrelevant to the energy, as commented above. However, the symmetry-breaking information may be largely affected, since $\bra{\phi_{k,+}(g_0)}\hat{\mathcal{O}}\ket{\phi_{k,-}(g_0)}\neq 0$, and thus $\bra{\psi(2\tau)}\hat{\mathcal{O}}\ket{\psi(2\tau)}$ will depend on the phases $\varphi_{k,p}$, and will be, in general, different from the initial condition  $\bra{\psi(2\tau)}\hat{\mathcal{O}}\ket{\psi(2\tau)}\neq \bra{\psi_0}\hat{\mathcal{O}}\ket{\psi_0}$. 

We can explicitly write the expression for the expectation value of $\hat{\mathcal{O}}$ after the completion of the cycle. Assuming that $\bra{\phi_{k,p}(g_0)}\hat{\mathcal{O}}\ket{\phi_{k,p}(g_0)}=0$ for $p=+,-$, then
\begin{align}
\bra{\psi(2\tau)}\hat{\mathcal{O}}\ket{\psi(2\tau)}=\sum_{k,k'}c_{k,+}c_{k',-}^*e^{-i(\varphi_{k,+}-\varphi_{k',-})}\bra{\phi_{k',-}(g_0)}\hat{\mathcal{O}}\ket{\phi_{k,+}(g_0)}+{\rm H.c}.
\end{align}

Now, since only the eigenstates with the same $k$ label and opposite parity are degenerate, we can further simplify the previous expression, and find that
\begin{align}
\bra{\psi(2\tau)}\hat{\mathcal{O}}\ket{\psi(2\tau)}\approx \sum_{k}c_{k,+}c_{k,-}^*e^{-i(\varphi_{k,+}-\varphi_{k,-})}\bra{\phi_{k,-}(g_0)}\hat{\mathcal{O}}\ket{\phi_{k,+}(g_0)}+{\rm H.c}.
\end{align}

If the initial state populates just a single doublet (e.g., only the coefficients $c_{0,+}$ and $c_{0,-}$ are non-zero), then the final value will be present an oscillatory behavior $\bra{\psi(2\tau)}\hat{\mathcal{O}}\ket{\psi(2\tau)}\propto \cos[\delta\varphi_{0}]\bra{\psi_0}\hat{\mathcal{O}}\ket{\psi_0}$, where $\delta\varphi_{0}=\varphi_{0,+}-\varphi_{0,-}=2\tau/(g_1-g_0)\int_{g_0}^{g_1} dg (E_{k,+}(g)-E_{k,-}(g))$ is the difference between the phases, depending on $\tau$. In this manner, tuning $\tau$ (assuming that the adiabatic approximation holds), the symmetry-breaking observable can be of any value between $\bra{\psi_0}\hat{\mathcal{O}}\ket{\psi_0}$ and $-\bra{\psi_0}\hat{\mathcal{O}}\ket{\psi_0}$. Hence, the information regarding the initial symmetry breaking is not scrambled, as the expectation value $\bra{\psi(2\tau)}\mathcal{O}\ket{\psi(2\tau)}$ may or may not reverse the initial value by simply tuning $\tau$. However, when the initial state populates a large number of eigenstates, and if the difference in the accumulated phases $\delta\varphi_{k}$ for different $k$s uniformly sample the interval $[0,2\pi)$, then, 
\begin{align}\label{eq:O_AQIS}
\bra{\psi(2\tau)}\hat{\mathcal{O}}\ket{\psi(2\tau)}\approx 0,
\end{align}
regardless of the specific time $\tau$. Note that the phase differences $\delta\varphi_{k}$ will only be non-zero if the system is driven from the symmetry-breaking to the normal phase, as $E_{k,+}=E_{k,-}$ otherwise, leading to $\delta\varphi_{k}=0$. Hence, the mechanism for AQIS requires $g_0$ to be in the symmetry-breaking phase and $g_1$ in the normal phase, thus forcing the system to traverse the excited-state quantum phase transition. 

A consequence of Equation~\eqref{eq:O_AQIS} is that the probability distribution of the operator $\hat{\mathcal{O}}$ for the final state $\ket{\psi(2\tau)}$ will exhibit a balanced distribution for $\gamma<0$ and $\gamma>0$, as if it were a symmetric state. In this manner, the initial symmetry-breaking information is scrambled in the quantum system due to the large number of populated eigenstates.   We can anticipate the expression given in Equation~\eqref{eq:O_AQIS}, i.e., that AQIS will be more effective the larger the support of the initial state over the eigenstates of $\hat{H}(g_0)$.  In the following, we illustrate  AQIS in the LMG and QRM models, supporting its effectiveness, robustness, and dependence on different initial states with numerical simulations. 

\section{Lipkin--Meshkov--Glick Model}\label{s:LMG}
\begin{figure}[h!]
    \centering
    \includegraphics[width=0.8\linewidth]{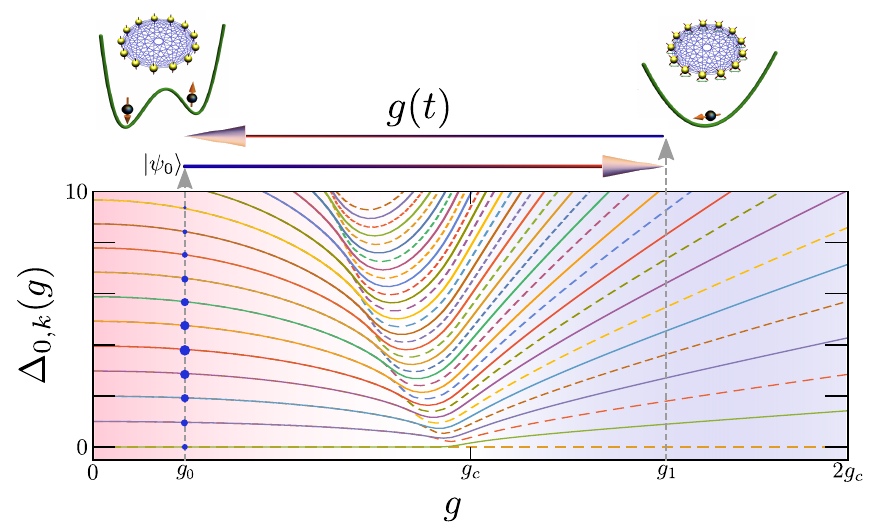}
    \caption{{\bf {Sketch of protocol that leads to an adiabatic scrambling of information}}. Information is encoded in an initially symmetry-breaking state $\lvert \psi_0 \rangle$, which populates a large collection of eigenstates in the symmetry-broken phase of the model at $g_0<g_c$. The system then undergoes a coherent adiabatic cycle $g(t)$, entering into the normal phase and back. The populations of the initial state remain unaltered, while the information is scrambled due to the change in the relative phases among degenerate eigenstates. The energy spectrum corresponds to the LMG for $N=300$ spins, where $\Delta_{0,k}(g)=E_{k}(g)-E_{0}(g)$, with the dashed and solid lines depicting the energies of even and odd parity eigenstates, respectively.}
    \label{fig1}
\end{figure}

The LMG model~\cite{Lipkin:65}, originally introduced in the context of nuclear physics, describes the long-range dipole--dipole interaction of $N$ spin-$\frac{1}{2}$ under a transverse magnetic field. This model exhibits QPT~\cite{Sachdev} at a certain critical value of the field strength $g_c$~\cite{Dusuel:04, Leyvraz:05, Vidal:07, Ribeiro:07, Ribeiro:08}, and has attracted renewed attention due to its experimental realization, e.g., with cold atoms~\cite{Zibold:10} or in trapped ions~\cite{Jurcevic:17}. Indeed, the LMG has served as a test bed for the exploration of different aspects of quantum critical systems~\cite{Caneva:08, Kwok:08, Yuan:12, Acevedo:14, Salvatori:14, Campbell:15, Puebla:15, Campbell:16, Defenu:18, Puebla:20, Mzaouali:21, Garbe:22, Gamito:22, Garbe:22b, Abah:22, Corps:22b, Santini:24}. The Hamiltonian of the $N$  spin-$\frac{1}{2}$ particles can be written as (where the unit of frequency is equal to $1$, 
 and $\hbar=1$)
\begin{align}
\hat{H}_{\rm LMG}=-g\hat{S}_z-\frac{\hat{S}_x^2}{N},
  \end{align}
where $\hat{S}_{\alpha}=\sum_{k=1}^N\hat{\sigma}_\alpha/2$ are the collective spin-{$1/2$} operators along the $\alpha$-direction ($\alpha=x,y,z$). Since the $\hat{H}(g)$ commutes with $\hat{S}^2=\hat{S}_x^2+\hat{S}_y^2+\hat{S}_z^2$, we restrict ourselves to the sector of maximum spin, i.e., $J=N/2$, and work in the Dicke basis, $\hat{S}_z\ket{J,M}=M\ket{J,M}$ with $-J\leq M\leq J$, and $\hat{S}^2\ket{J,M}=J(J+1)\ket{J,M}$. For simplicity, we will consider $N$ to be even. In the $N\rightarrow \infty$, the previous Hamiltonian features QPT at the critical value $g_c=1$~\cite{Ribeiro:08, Ribeiro:07}, which divides the spectrum into two phases (see Figure~\ref{fig1} for a representation of the energy spectrum for $N=300$ spins). For $g<g_c$, there is a symmetry-breaking $\mathbb{Z}_2$ phase up to critical energy $E_c(g)$, dependent on $g$, where the eigenstates of different parity are degenerate; to the contrary, for energies above $E_c(g)$ or for $g>g_c$, there is a normal phase where the parity is well defined and the eigenstates are no longer degenerate. The location of the critical excitation energy $E_c(g)$ corresponds to the so-called excited-state quantum phase transition (ESQPT), whose main hallmark consists of a logarithmically diverging density of states at $E_c(g)$~\cite{Leyvraz:05, Puebla:13b, Cejnar:06, Cejnar:08, Caprio:08, Brandes:13, Puebla:16, PerezFernandez:17, Stransky:14, Stransky:15}. In the LMG model, the spin operator $\hat{S}_x$ plays the role of the operator $\hat{\mathcal{O}}$ introduced in Section~\ref{s:AQIS}, whose ground-state expectation value serves as a good order parameter for the ground-state QPT. 

\begin{figure}[h!]
    \centering
    \includegraphics[width=0.8\linewidth]{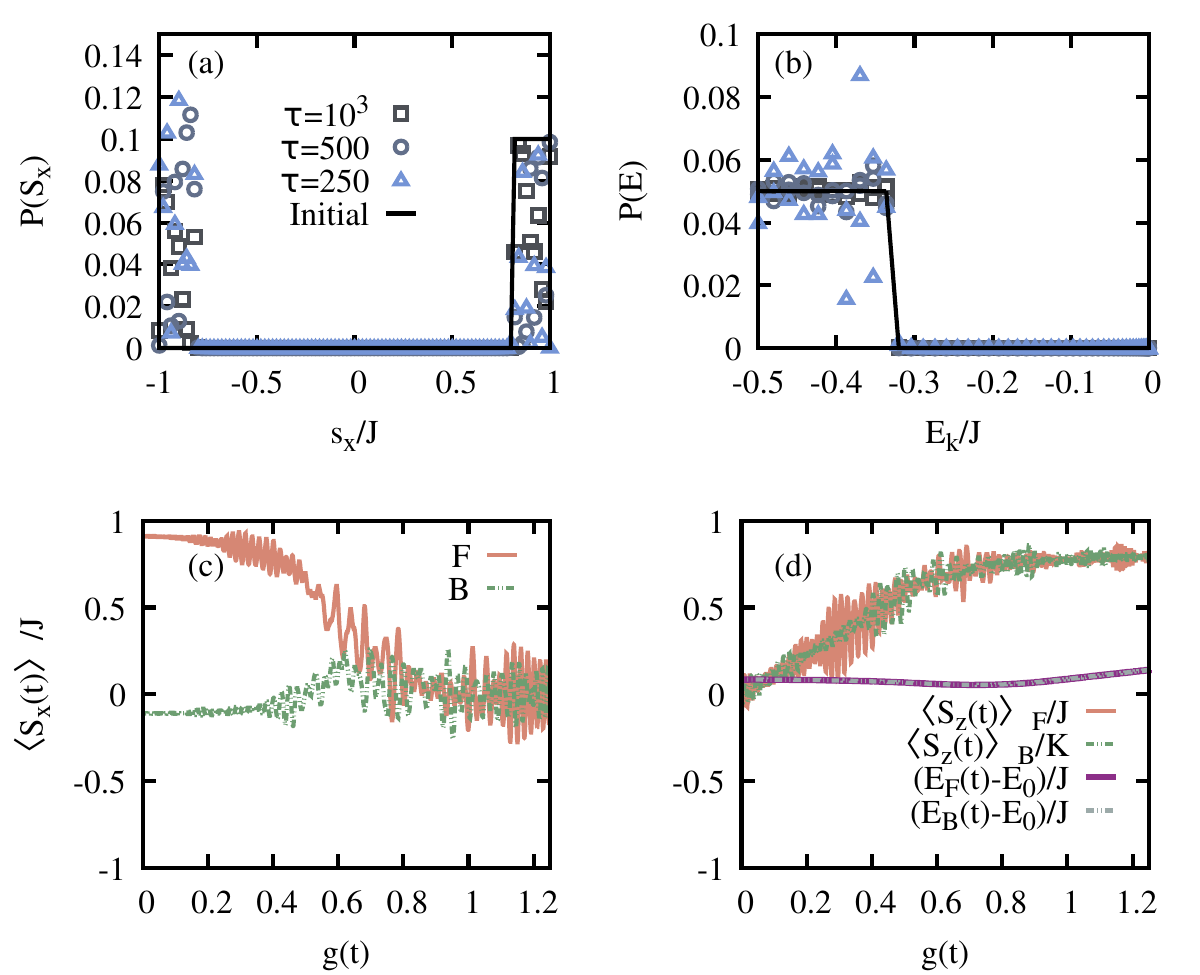}
    \caption{ {\bf {Adiabatic quantum scrambling of the symmetry-breaking information in the LMG model.}} {Panel}  (\textbf{a}) {shows} 
 the initial and final probability distribution over the observable $\hat{S}_x$ for an initial microcanonical-like state undergoing a fully coherent cycle from the symmetry-breaking to the normal phase and back. {Panel} (\textbf{b}) shows the energy probability distribution $P(E)$ for the initial and final states after the cycle (same $\tau$ values as in (\textbf{a})). {Panel} (\textbf{c}) illustrates the dynamics of the expectation value $\bra{\psi(t)}\hat{S}_x\ket{\psi(t)}$ as a function of the time-dependent control $g(t)$ for $\tau=500$. Here, $F$ ($B$) denotes the forward (backward) process from $g_0\rightarrow g_1$ ($g_1\rightarrow g_0$). Note that $\langle S_x\rangle\approx 0$ after the adiabatic protocol. In (\textbf{d}) the dynamics of other relevant observables not related to the symmetry-breaking information are shown, and thus not scrambled during the adiabatic protocol. The employed parameters are $N=100$ spins, $g_0=0$, $g_1=1.25$, $N_{\rm mc}=10$. }%MDPI: please check if all ``Pane''
    \label{fig2}
\end{figure}

In the following, we illustrate how the mechanism described in Section~\ref{s:AQIS} applies to the LMG model, i.e., the adiabatic quantum information scrambling of the symmetry-breaking initial state. 
To study the effectiveness of AQIS as a function of the number of populated eigenstates, i.e., the validity of Equation~\eqref{eq:O_AQIS}, we first consider an initial microcanonical-like state that uniformly populates the first $2N_{\rm mc}$ eigenstates. That is, 
\begin{align}\label{eq:psi_mc}
\ket{\psi_0}=\frac{1}{\sqrt{2N_{\rm mc}}}\sum_{k=0}^{N_{\rm mc}-1}\left(\ket{\phi_{k,+}(g_0)}+x_k\ket{\phi_{k,-}(g_0)}\right),
\end{align}
where the coefficients $x_k=-1,+1$ are chosen according to the sign of $\langle \phi_{k,+}| \hat{S}_x| \phi_{k,-} \rangle$, so that $\bra{\psi_0}\hat{S}_x\ket{\psi_0}$ is positive, and the probability distribution $P(S_x)$ only shows non-zero values in the positive branch.
See Figure~\ref{fig2}a,b for the initial probability distribution of this microcanonical-like state where $g_0=0$ and $N=100$ spins with $N_{\rm mc}=10$. The dynamical protocol follows Equation~\eqref{eq:gt} with $g_1=1.25$ to ensure that the system is completely driven into the normal phase (cf. Figure~\ref{fig1}). The final probability distributions are also plotted in Figure~\ref{fig2}a,b for the $\hat{S}_x$ and energy, respectively. Considering slow ramps ($\tau\gg 1$), we first note that the energy probability distribution $P(E)$ remains unaltered, i.e., the coefficients $c_{k,p}$ approximately hold constant. To the contrary, $P(S_x)$ significantly differs from the initial distribution as a consequence of  AQIS. The final state $\ket{\psi(2\tau)}$ populates both branches of the observable $\hat{S}_x$, and thus its expectation value after the completion of the cycle vanishes, $\bra{\psi(2\tau)}\hat{S}_x\ket{\psi(2\tau)}\approx 0$. This can be better visualized in Figure~\ref{fig2}c, where we show the evolution of $\bra{\psi(t)}\hat{S}_x\ket{\psi(t)}$ as a function of $g(t)$. Since the dynamical evolution is to a very good approximation adiabatic, other observables not related to the symmetry-breaking information remain unchanged, such as $\hat{S}_z$ or the energy (cf. Figure~\ref{fig2}d). In the following, and to ease the numerical simulations, we assume that the evolution is adiabatic, and thus make use of the approximation given in Equation~\eqref{eq:PsiAd} to compute the final state after the cycle. 

\subsection{Effectiveness of the Quantum Information Scrambling}
\begin{figure}[h!]
    \centering
    \includegraphics[width=0.8\linewidth]{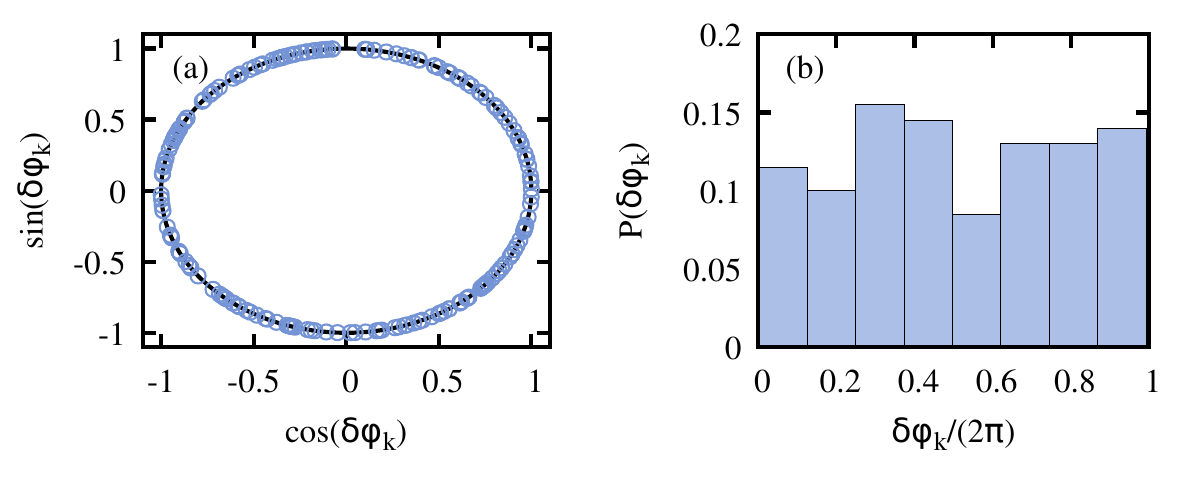}
    \caption{{\bf {Distribution of the phases} $\delta\varphi_k$.} {Panel} %
 (\textbf{a}) shows the obtained phases plotted in a unit circle, while the distribution of the phases in the range $[0,2\pi)$ is shown in (\textbf{b}), revealing an approximately uniform distribution. The results correspond to the LMG model with $N=2000$ spins and for the first 200 eigenstates, $\tau=10^3$ with $g_0=0$ and $g_1=1.25$.}
    \label{fig3}
\end{figure}

The effectiveness of the AQIS of the symmetry-breaking information is intimately related to the distribution of the phases $\delta\varphi_k$, i.e., the difference of the accumulated phase among eigenstates of opposite parity, as discussed in Section~\ref{s:AQIS}. Thus, the proper quantum information scrambling requires $\delta\varphi_k$ to uniformly sample $[0,2\pi)$. In Figure~\ref{fig3}, we show the resulting distribution of these phases $\delta\varphi_k$ for a LMG model comprising $N=2000$ spins for the first 200 eigenstates, and $\tau=10^3$ with $g_0=0$ and $g_1=1.25$. Note that it approximately corresponds to a uniform distribution in the range $[0,2\pi)$. Similar results can be found for other choices of $\tau$, $N$, $g_0$, and $g_1$, provided that $g_1$ ensures that the system enters in the normal phase. Otherwise, if the system remains in the symmetry-breaking phase during the whole cycle, then $\delta\varphi_k=0 \ \forall k$, since $E_{k,+}(g)=E_{k,-}(g)$ during the protocol, and information scrambling will not occur.
\unskip

\begin{figure}[h!]
    \centering
    \includegraphics[width=0.8\linewidth]{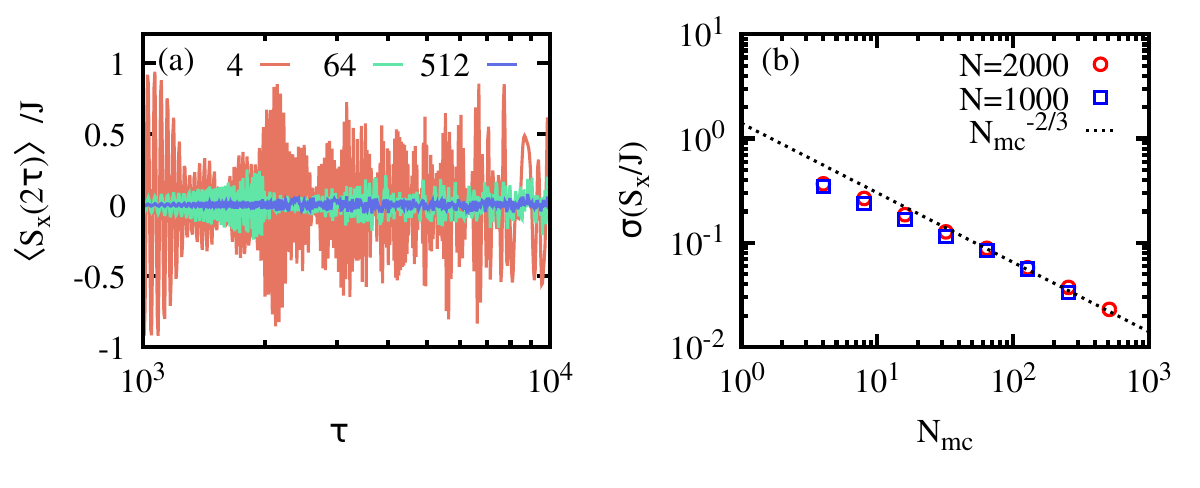}
    \caption{{\bf {Effectiveness of  AQIS as a function of the number of populated eigenstates}. Panel 
 (\textbf{a})~shows the final expectation value of $\hat{S}_x$, i.e., $\bra{\psi(2\tau)}\hat{S}_x\ket{\psi(2\tau)}$ as a function of $\tau$, ensuring the adiabatic approximation, and for a different number of populated eigenstates $N_{\rm mc}$, i.e., $4$, $64$ and $512$. Panel (\textbf{b})~shows the the resulting standard deviation of the results in (\textbf{a}) for their average value, as a function of $N_{\rm mc}$. See the main text for further details.}}
    \label{fig4}
\end{figure}

As stated in Section~\ref{s:AQIS}, the effectiveness of AQIS depends on the number of populated eigenstates in the initial state, i.e., on the support of $\ket{\psi_0}$ on the eigenstates of $\hat{H}(g_0)$. The initial state given in Equation~\eqref{eq:psi_mc} allows us to systematically analyze the validity of Equation~\eqref{eq:O_AQIS} as $N_{\rm mc}$ increases. We first note that, if $N_{\rm mc}\approx 1$, then the final expectation value of $\hat{S}_x$ will simply undergo an oscillatory dependent on $\tau$. This can be seen in Figure~\ref{fig4}a for an initial state with $N_{\rm mc}=4$, which hardly scrambles the symmetry-breaking information, as one can recover the initial value $\bra{\psi_0}\hat{S}_x\ket{\psi_0}$ at a suitable time $\tau$. Yet, as $N_{\rm mc}$ grows, this is no longer the case, due to the uniform distribution of the phases $\delta\varphi_k$, which forces the final expectation value to average to zero, i.e., $\bra{\psi(2\tau)}\hat{S}_x\ket{\psi(2\tau)}\approx 0$. The effectiveness of AQIS can therefore be quantified in terms of the fluctuations of $\bra{\psi(2\tau)}\hat{S}_x\ket{\psi(2\tau)}$ around its average value, i.e., 
\begin{align}
    \sigma^2(S_x)=\overline{\left( \bra{\psi(2\tau)}\hat{S}_x\ket{\psi(2\tau)}-\overline{\bra{\psi(2\tau)}\hat{S}_x\ket{\psi(2\tau)}}\right)^2}.
\end{align}

In the previous expression, $\overline{X}$ corresponds to the average value of $X$ in an interval of quench times $\tau\in[\tau_0,\tau_1]$, where $\tau_0$ already ensures the validity of the adiabatic approximation. If the information is perfectly scrambled, then $\sigma^2(S_x)=0$, so that the adiabatically evolved final state features $\bra{\psi(2\tau)}\hat{S}_x\ket{\psi(2\tau)}=0$ independently of $\tau$. Therefore, the smaller the $\sigma(S_x)$, the more effective AQIS is. The results are plotted in Figure~\ref{fig4}b, obtained for $\tau_0=10^3$, $\tau_1=10^4$, and similar parameters as in the previous figures, namely $g_0=0$, $g_1=1.25$, and for two system sizes, $N=1000$ and $2000$ spins. As the number of populated states grows, $N_{\rm mc}$, the variance $\sigma^2(S_x)$ decreases, thus indicating a better scrambling performance. A fit to the numerical results reveals a dependence $\sigma(S_x)\propto N_{\rm mc}^{-2/3}$ (cf. Figure~\ref{fig4}b). 

\subsection{Loschmidt Echo and Out-of-Time-Ordered Correlator}
The previous results support the effectiveness of  AQIS of the symmetry-breaking information of the initial state $\ket{\psi_0}$. However, since the state evolves in a fully coherent manner, $\ket{\psi(2\tau)}=\hat{U}(2\tau)\ket{\psi_0}$ where $\hat{U}(t)=\hat{\mathcal{T}}e^{-i\int_0^{t} dt' \hat{H}(t')}$ corresponds to unitary time-evolution operator from $0$ to $t$ under the protocol $g(t)$, the initial state can be recovered by a perfect time-reversal operation, i.e., $\ket{\psi_0}=\hat{U}(2\tau)^\dagger \ket{\psi(2\tau)}$. This corresponds to an evolution of the final state under $\hat{H}(g)\rightarrow -\hat{H}(g) \ \forall g$. If such a time-reversal operation is perfect, then the symmetry-breaking information can be recovered from the scrambled state $\ket{\psi(2\tau)}$. However, as is customary in closed systems, the retrieval of the initial information highly depends on any potential deviation from a perfect time-reversal evolution, and thus any small mismatch will hinder the recovery of the initial state. This motivates the analysis of the Loschmidt echo,
\begin{align}
L(\tau,\delta t)=\big\lvert\langle \psi_0 | \hat{U}^\dagger(2\tau+\delta t)\hat{U}(2\tau)|\psi_0 \rangle \big\rvert,
  \end{align}
  which quantifies the overlap between the final state $\ket{\psi(2\tau)}=\hat{U}(2\tau)\ket{\psi_0}$ and the potentially restored initial state upon a time-reversal evolution with a small time-delay $\delta t$. Hence, for $\delta t=0$, it follows that $L(\tau,\delta t=0)=1 \ \forall \tau$. The scrambling due to AQIS will be robust the smaller the Loschmidt echo for small time-delays, i.e., if $L(\tau,\delta t)\approx 0$ for $\delta t\ll 1$.

\begin{figure}[h!]
    \centering
    \includegraphics[width=0.7\linewidth]{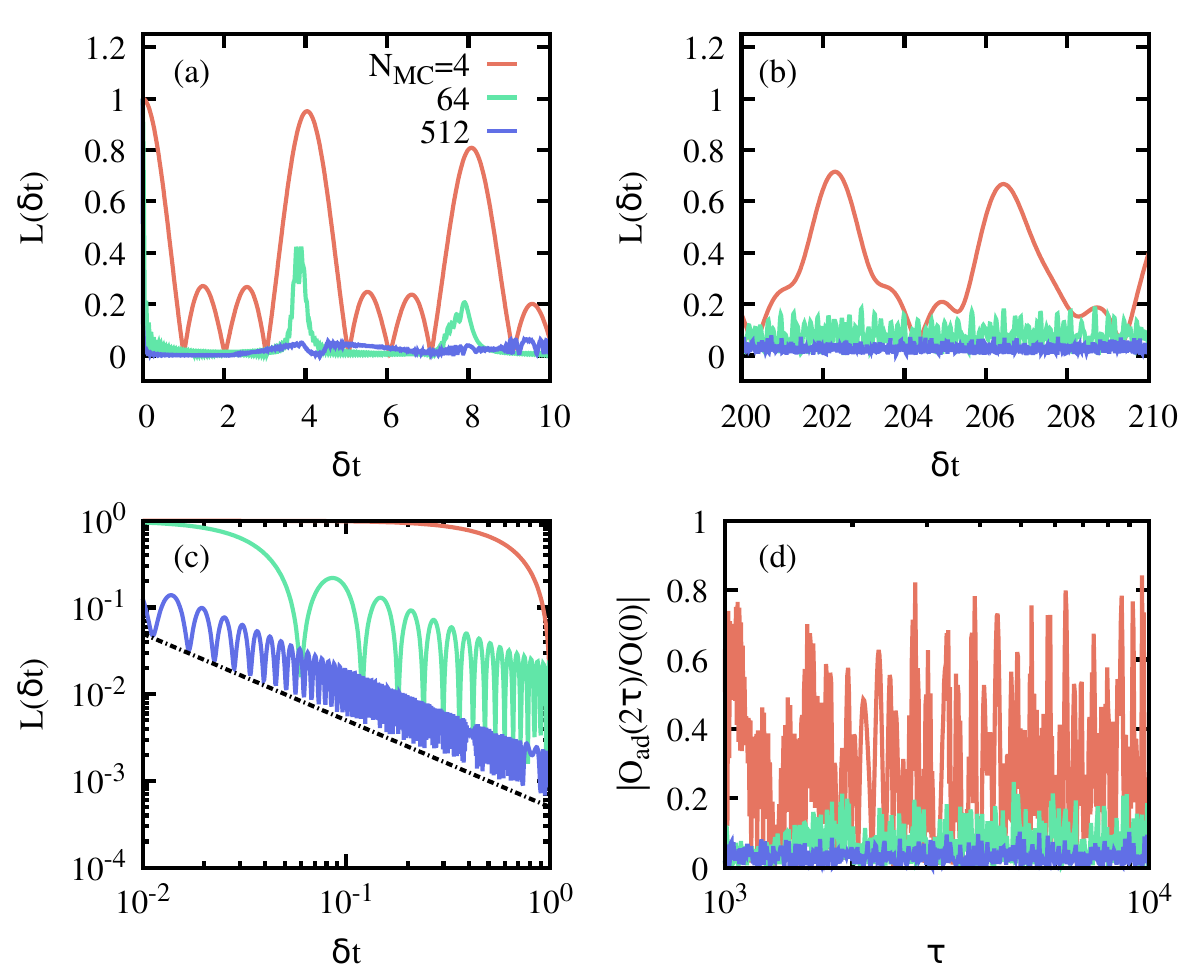}
    \caption{{\bf {Loschmidt echo and OTOC for the LMG model}}. {Panels}
 (\textbf{a}--\textbf{c}) show the Loschmidt echo $L(\delta t)$ as a function of the time mismatch $\delta t$ and for initial states with different populated eigenstates $N_{\rm mc}$. Note strong revivals even at late times (\textbf{b}), and the decaying echo $1/\delta t$ (dashed line) in (\textbf{c})~when $N_{\rm mc}\gg 1$. Panel (\textbf{d}) corresponds to a rescaled OTOC, which signals a similar behavior to the expectation value of $S_x$. The simulation parameters are $N=2000$ spins, $g_0=0$, and $g_1=1.25$. See the main text for further details. }
    \label{fig5}
\end{figure}

  As we are interested in an adiabatic evolution, i.e., for sufficiently long $\tau$, we rely again on the adiabatic approximation. It is easy to see that, under this approximation, one obtains the simple expression
\begin{align}
L(\tau,\delta t)=\Bigg\lvert\sum_{\substack{k=0\\ p=\pm}} |c_{k,p}|^2e^{i\delta t \varphi_{k,p}}\Bigg\rvert, 
  \end{align}
with $\sum_{k,p} |c_{k,p}|^2=1$. Note that under the adiabatic approximation, the previous quantity does not depend on $\tau$, i.e., $L(\tau,\delta t)\equiv L(\delta t)$. Again, as for the effectiveness of AQIS (Equation~\eqref{eq:O_AQIS}), for initial states with a large support  and uniformly distributed phases $\varphi_{k,p}$, one expects $L(\delta t)\approx 0$, even for $0<\delta t\ll 1$. The results are plotted in Figure~\ref{fig5}a--c, which precisely reveal a vanishingly small echo $L(\delta t)\approx 0$ for small time mismatches $\delta t$. Although not explicitly shown, we note that similar results can be obtained if, instead of a time mismatch $\delta t$, a deviation in another quantity is considered,  such as in $g_{0,1}$. In particular, we observe a decay $L(\delta t)\propto 1/\delta t$ for short $\delta t$ and $N_{\rm mc}\gg 1$. As expected, in states with little support %EE: Please check intended meaning has been retained.
 (few populated eigenstates), $L(\delta t)$ displays large revivals, and never decays to $0$, a clear indication of the failure of quantum information scrambling.  
As commented in Section~\ref{s:intro}, OTOCs have been proven a valuable tool for studying information scrambling. In a sudden quench scenario, scrambling is related to the short-time behavior of the OTOC for a relevant observable~\cite{Alba:19}. In our case, we can define the OTOC as
\begin{equation}\label{eq:otoc}
\begin{split}
    O(t)&=\langle \psi_0|\hat{S}_x(t) \hat{S}_x \hat{S}_x(t) \hat{S}_x|\psi_0\rangle,\\
    &=\langle \psi_0|\hat{U}^\dagger(t) \hat{S}_x \hat{U}(t) \hat{S}_x \hat{U}^\dagger(t)\hat{S}_x \hat{U}(t) \hat{S}_x|\psi_0\rangle.
\end{split}
\end{equation}

Since  AQIS is effective only after the cycle has been completed, we focus on $O(2\tau)$. Again, relying on the adiabatic approximation of the evolved state, we approximate $\hat{U}(2\tau)$ by its adiabatic counterpart $\hat{U}^{\rm Ad}(2\tau)$, so that the populations in the eigenbasis of the instantaneous Hamiltonian remain constant, and the only change is due to the accumulated phase during the cycle. In this manner, we can compute the adiabatic OTOC, denoted here as $O^{\rm Ad}(2\tau)$ and fulfilling $O^{\rm Ad}(2\tau)\approx O(2\tau)$ for $\tau\gg 1$. The results for $O^{\rm Ad}(2\tau)$, plotted in Figure~\ref{fig5}d, reveal essentially the same behavior as $\bra{\psi(2\tau)} \hat{S}_x\ket{\psi(2\tau)}$. That is, the adiabatic dynamics scramble the symmetry-breaking information contained in the initial state, and thus $O(2\tau)\approx 0$. This becomes more effective the more eigenstates are significantly populated in the initial state. Conversely, for small support (e.g., $N_{\rm mc}=4$ corresponding to the red line in Figure~\ref{fig5}d), the OTOC features large oscillations, potentially reaching its initial value. %\FG
{Although not explicitly shown here, it is worth mentioning that the corresponding OTOC for operators not related to the symmetry-breaking information, such as $\hat{S}_z$, do not show scrambling. As a consequence, these OTOCs are approximately constant, and close to their initial value.}

\subsection{Symmetry-Breaking Thermal States}
Having systematically analyzed the mechanism of AQIS in the LMG model and its effectiveness in terms of the number of populated eigenstates, we turn now our attention to a different initial state. Moreover, we will study how the reported AQIS depends on the control parameter $g_1$, i.e., on whether the system is driven from the symmetry-breaking into the normal phase or not. 

\begin{figure}[h!]
\centering
    \includegraphics[width=0.7\linewidth]{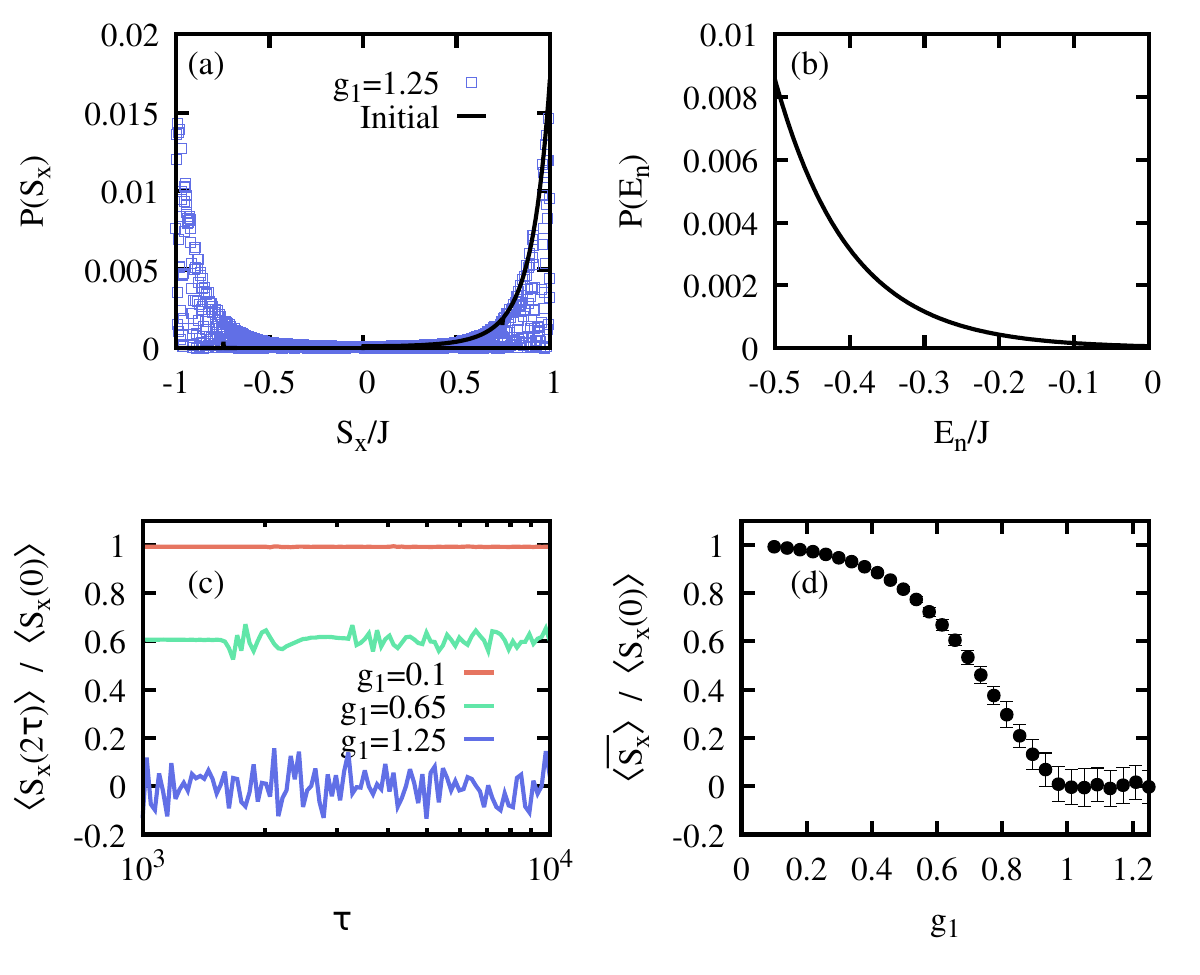}
\caption{{\bf {AQIS for a symmetry-breaking thermal state in the LMG model.}} (\textbf{a}) {Initial} 
 probability distribution $P(S_x)$ over the observable $\hat{S}_x$ for $\hat{\rho}_{\beta, SB}$ with $\beta=0.02$ for $N=10^3$ and $g_0=0$ (solid black line), together with the probability for the state upon an adiabatic protocol for $\tau=10^3$ and $g_1=1.25$ (blue squares), which equally populates both $S_x$ branches.  (\textbf{b}) Population of the initial eigenstate for the symmetry-breaking thermal state $\hat{\rho}_{\beta,SB}$, as in (\textbf{a}). (\textbf{c}) Final expectation value of $\hat{S}_x$ after an adiabatic protocol from $g_0=0$ to $g_1=0.1$ (red), $0.65$ (green), and $1.25$ (blue), as a function of $\tau$ and for $\hat{\rho}_{\beta, SB}$ with $\beta=0.02$ and $N=10^3$, as in (\textbf{a},\textbf{b}). For $g_1=0.1$, the system is not driven into the normal phase. For $g=0.65$, the state enters partially, while for $g_1=1.25$ it is driven completely into the normal phase, leading to a full AQIS. (\textbf{d}) Average expectation value of $\hat{S}_x$ in the range $\tau\in\{10^3,10^4\}$ as a function of the final coupling strength $g_1$, revealing an order-parameter-like behavior. The error bars correspond to a standard deviation for the average of $\langle \hat{S}_x(2\tau)\rangle$. See the main text for details.}
\label{fig6}
\end{figure}

Here, we consider a different initial state, namely, a symmetry-breaking thermal state with inverse temperature $\beta$ ($k_b=1$),
\begin{align}\label{eq:rhoSB}
    \hat{\rho}_{\beta,SB}=\frac{1}{Z}\sum_{k}e^{-\beta E_k(g_0)}\ket{\psi_{k,SB}(g_0)}\bra{\psi_{k,SB}(g_0)},
\end{align}
where $Z$ is the partition function that ensures a proper normalization of the state, ${\rm Tr}[\hat{\rho}_{\beta,SB}]=1$, while $\ket{\psi_{k,SB}(g_0)}=\frac{1}{\sqrt{2}}(\ket{\psi_{k,+}(g_0)}+x_k\ket{\psi_{k,-}(g_0)})$ and $x_k=\pm 1$, depending on the expectation value of $\hat{S}_x$. As for Equation~\eqref{eq:psi_mc}, we consider a maximally symmetry-breaking state with ${\rm Tr}[\hat{\rho}_{\beta,SB}\hat{S}_x]>0$ so that $P(S_x)$ shows only non-zero probability for positive eigenvalues of $\hat{S}_x$ (cf. Figure~\ref{fig6}a), while $P(E)$ displays the standard exponentially decaying populations (cf. Figure~\ref{fig6}b). Hence, the state $\hat{\rho}_{\beta,SB}$ corresponds to a thermal state projected onto one of the two symmetry-breaking branches, i.e., to a generalized Gibbs state. 

We then analyze how the realization of the adiabatic cycle scrambles the initial information, depending on the value of $g_1$. As shown in Figure~\ref{fig1}, to ensure that the whole state is driven into the normal phase, $g_1>1$. For a final value $g_1<1$, only a fraction of excited states will enter the normal phase, thus rendering  AQIS ineffective, as $\delta\varphi_k=0$ for those eigenstates that remain within the symmetry-breaking phase. The results are plotted in Figure~\ref{fig6}. As for the initial state given in Equation~\eqref{eq:psi_mc}, the final distribution $P(S_x)$ largely differs from the initial one, since both branches are populated, ensured by $g_1=1.25$. For other choices of $g_1$, the final $P(S_x)$ may fail to be a balanced distribution. This is shown in Figure~\ref{fig6}(c), where the final expectation value $\langle \hat{S}_x(2\tau)\rangle={\rm Tr}[\hat{U}(2\tau)\hat{\rho}_{\beta,SB}\hat{U}^\dagger(2\tau)S_x]$ is plotted as a function of $\tau$, and for three different values of $g_1$. For $g_1=0.1$ the system remains in the symmetry-breaking phase during the whole cycle, and thus $\delta\varphi_k=0 \ \forall k$, resulting in the absence of scrambling. For $g_1=0.65$, only a fraction of eigenstates are driven into the normal phase, partially scrambling the initial information. Finally, for $g_1=1.25$, the whole system enters in the normal phase, which corresponds to  AQIS, as discussed previously. To better visualize this effect, in Figure~\ref{fig6}(d), we show the average value of $\langle S_x(2\tau)\rangle$ in the range $\tau\in[10^3,10^4]$ as a function of $g_1$, which behaves as an order parameter for  AQIS. 
\unskip
\begin{figure}[h!]
\centering
    \includegraphics[width=\linewidth]{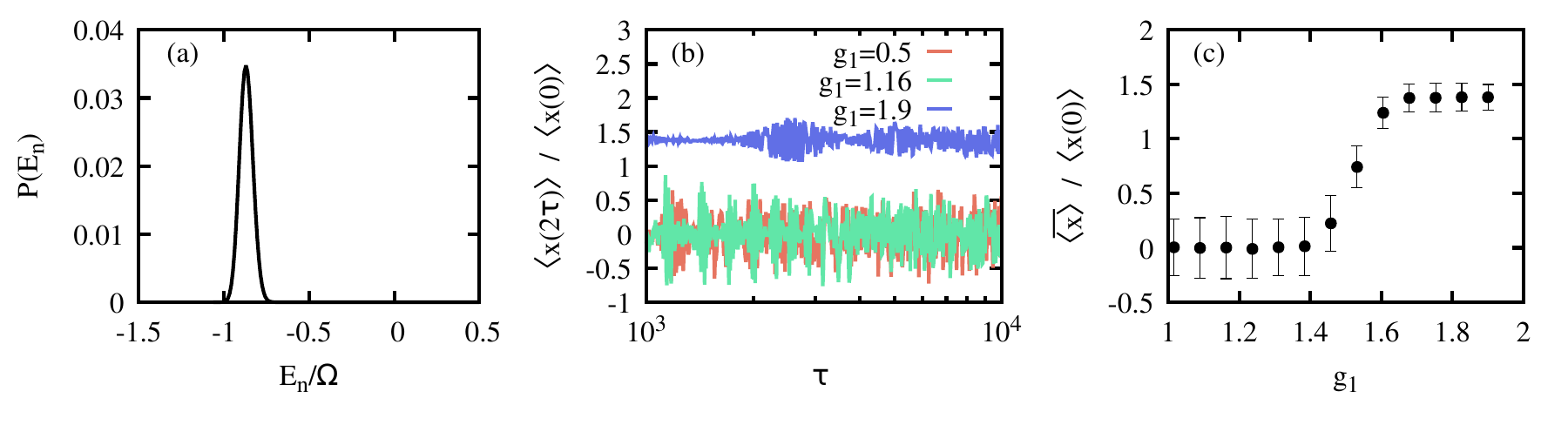}
\caption{{\bf {AQIS for the QRM}}. (\textbf{a}) {Initial}
 energy probability distribution $P(E_n)$ for $\ket{\psi(0)}=\ket{\alpha}_m\ket{0}_s$ for $\Omega=100\omega$, initial rescaled coupling $g_0=2$, and $\alpha=5$. The normal phase finds itself above the critical excitation energy $E_c/\Omega=-1/2$. Upon the adiabatic protocol, the energy probability distribution is equivalent to the initial one. (\textbf{b}) Expectation value of the symmetry-breaking observable $\hat{x}$ after the adiabatic protocol of duration $\tau$ from $g_0=2$ to $g_1$, as indicated in the key, assuming the adiabatic approximation.  (\textbf{c}) Average symmetry-breaking expectation value, in the range $\tau\in\{10^3,10^4\}$, as a function of the final coupling $g_1$. For $g_1\lesssim 1.5$, the whole state enters in the normal phase, leading into a complete adiabatic information scrambling. Error bars correspond to the standard deviation for $\langle x(2\tau)\rangle$, as plotted in (\textbf{b}). See the main text for further details. }
\label{fig7}
\end{figure}

\section{Quantum Rabi Model}\label{s:QRM}
The QRM describes the coherent interaction between a spin-$1/2$ and a single bosonic mode, and thus constitutes a fundamental model in the realm of light-matter interaction and of key relevance in quantum technologies. The Hamiltonian of the QRM can be written as\vspace{6pt}
\begin{align}
    \hat{H}_{\rm QRM}(\lambda)=\frac{\Omega}{2}\hat{\sigma}_z+\omega \hat{a}^\dagger \hat{a}+\lambda (\hat{a}+\hat{a}^\dagger)\hat{\sigma}_x,
\end{align}
where $\Omega$, $\omega$, and $\lambda$ correspond to the frequency of the qubit, bosonic mode, and interaction strength among them. Despite comprising just a single spin, this model has been shown to display ground-, excited-state, and even dynamical quantum phase transitions in the limit of $\Omega/\omega\rightarrow \infty$~\cite{Bakemeier:12, Hwang:15, Puebla:16, Puebla:20b}. In this particular parameter limit, which plays the role of the standard thermodynamic limit where critical phenomena typically take place, the ground-state QPT occurs at $\lambda_c = \sqrt{\omega\Omega}/2$. Hence, for convenience, we introduce the rescaled interaction strength, $g=\lambda/\lambda_c$, so that $g_c=1$, as in the LMG. 

The phase diagram of the critical QRM is divided into three regions. For $g<g_c$, one has a normal phase. For $g>g_c$, and energy $E_k(g)<E_c=-\Omega/2$, one finds the symmetry-breaking phase where the $\mathbb{Z}_2$ parity symmetry of the total number of excitations $\hat{\Pi}$ is spontaneously broken. Finally, for $g>g_c$ and $E_k(g)>E_c$, the symmetry is restored, and the model is again in the normal phase. The excited-state QPT takes place at the critical energy $E_c=-\Omega/2$~\cite{Puebla:16}.  Besides the microscopic details, the universal critical features are equivalent to those of the LMG model, with a similar energy spectrum as the one depicted in Figure~\ref{fig1} (note, however, the difference in the location of the normal ($g<g_c$) and symmetry-breaking phases ($g>g_c$)). 

In the following, we exemplify that  AQIS also applies to the QRM. For that, we choose  
\begin{align}
    \ket{\psi_0}=\ket{\alpha}_m\ket{0}_s,
\end{align}
as our initial symmetry-breaking state. There, $\ket{\alpha}_m=\hat{D}(\alpha)\ket{0}_m=\exp \pas{\alpha \hat{a}^\dagger-\alpha^*\hat{a}}\ket{0}_m$ represents a coherent state with amplitude $\alpha$, and $\ket{0}_m$ is the vacuum state, while the spin is its ground state, $\hat{\sigma}_z\ket{0}_s=-\ket{0}_s$. This corresponds to a state in the other well %EE: Please check intended meaning has been retained.
 of the effective double-well potential in the symmetry-breaking phase~\cite{Hwang:15,Puebla:16}. The symmetry-breaking information is then attributed to the expectation value of the position operator, $\hat{x}=(\hat{a}^\dagger+\hat{a})/\sqrt{2}$.

The adiabatic protocol is performed from $g_0=2$ to $g_1$, according to Equation~\eqref{eq:gt}. Relying on the adiabatic approximation, and choosing $\alpha=5$ for a critical QRM with $\Omega=100\omega$, we compute the final state, and obtain $\bra{\psi(2\tau)}\hat{x}\ket{\psi(2\tau)}$ as a function of $g_1$. Numerical simulations have been performed with  $1000$ Fock states.  The results are gathered in Figure~\ref{fig7}. First, the energy probability distribution is shown in Figure~\ref{fig7}a, indicating that the state is contained within the symmetry-breaking phase, since all of the populated eigenstates have energies below the critical one, $E_c=-\Omega/2$. As before,  AQIS becomes effective when the whole state is driven into the normal phase. For the considered initial state, this happens for $g_1\lesssim 1.5$ (cf. Figure~\ref{fig7}b,c). Note that this corresponds to a larger parameter than its critical value $g_c=1$, where the QPT takes place. Instead, the value $g\approx 1.5$ marks the position when the energy of the adiabatically driven state surpasses the critical energy at which an excited-state QPT occurs~\cite{Puebla:16}. 

\section{Conclusions}\label{s:conc}
In this article, we have analyzed the adiabatic scrambling of the symmetry-breaking information encoded in an initial state. The mechanism for such adiabatic quantum information scrambling is detailed, which is related to the difference in accumulated phases among degenerated eigenstates due to the driving from a symmetry-breaking phase into a normal phase, and thus to phase transitions taking place both in the ground and excited states of the system. Owing to its adiabatic nature, the protocol does not alter either the energy or the expectation values of observables independent of the symmetry-breaking information. To the contrary, as a consequence of the scrambling, the expectation value of an operator that quantifies the symmetry-breaking becomes approximately zero, provided that the initial state has large support in the eigenstates of the Hamiltonian, and the relative difference among the accumulated phases of eigenstates of opposite parity are uniformly distributed. We showcase the adiabatic quantum information scrambling in the Lipkin--Meshkov--Glick and quantum Rabi models, and quantify the effectiveness of the scrambling for different initial states. The potential information retrieval is studied utilizing the Loschmidt echo, which indicates that an effective scrambling would require an almost perfect time-reversal protocol to restore the initial information. Our results demonstrate the intriguing interplay between quantum information, many-body systems, and critical phenomena.  

\vspace{6pt} 
\section*{Author Contributions:}
Conceptualization, R.P. and F.J.G.-R.; Formal analysis, R.P. and F.J.G.-R.; Investigation, R.P. and F.J.G.-R.; Writing – original draft, R.P. and F.J.G.-R. All authors have read and agreed to the published version of the manuscript.
\section*{Funding:}
F.J.G- R is grateful for financial support from the Spanish MCIN with funding from the European Union Next Generation EU (PRTRC17.I1) and the Consejer\'ia de Educaci\'on from Junta de Castilla y Leon through the QCAYLE project, as well as grants PID2023-148409NB-I00 MTM funded by AEI/10.13039/501100011033, and RED2022-134301-T.

\section*{Data Availability Statement:}
The raw data supporting the conclusions of this article will be made available by the authors on request.
\section*{Conflicts of Interest:} 
The authors declare no conflicts of interest.
\bibliography{Mybib}{}
\bibliographystyle{apsrev4-1}

\end{document}